# Anomalous friction in suspended graphene[1]

A. Smolyanitsky* and J.P. Killgore

Materials Reliability Division, National Institute of Standards and Technology, Boulder, CO 80305

*corresponding author: alex.smolyanitsky@nist.gov

**Abstract**

Since the discovery of the Amonton's law and with support of modern tribological models, friction between surfaces of three-dimensional materials is known to generally increase when the surfaces are in closer contact. Here, using molecular dynamics simulations of friction force microscopy on suspended graphene, we demonstrate an *increase of friction when the scanning tip is retracted away* from the sample. We explain the observed behavior and address why this phenomenon has not been observed for isotropic 3-D materials.

**Introduction**

Frictional properties of atomically thin layers have recently been studied experimentally [1-3] and theoretically [4-6]. A low (0.001 to 0.004) friction coefficient of suspended graphene has been reported [1], as well as an intriguing dependence of friction on the number of stacked layers [2, 4, 6]. Despite graphene's suggested promise as a revolutionary solid-state lubricant, our knowledge of the frictional properties of atomically thin layers is far from complete.

---

[1]Contribution of the National Institute of Standards and Technology, an agency of the US government. Not subject to copyright in the USA.

Often, friction force is estimated from the proportionality of contact area to load predicted in contact-mechanical models. Models, such as the Johnson-Kendall-Roberts (JKR) [7], and Derjaguin-Müller-Toporov (DMT) [8] are continuum-based extensions of the Hertzian contact theory [9]. While generally successful in explaining friction between smooth surfaces of various geometries, these models are inadequate in predicting the full effect of surface deformation, which is especially important for atomically thin surfaces in a sliding contact at the nanoscale [2, 6]. A serious limitation of the continuum models is the lack of atomistic insight into the tip-sample interactions. In response, atomistic potential based continuum theories [10] and detailed atomistic simulations have emerged as methods to describe the multitude of effects contributing to friction [5, 6, 11-13]. Here, we present the results of atomistic molecular dynamics (MD) simulations of friction force microscopy (FFM) scans of suspended graphene. Scans were simulated in repulsive and adhesive regimes to reveal the nature of the friction force versus normal force relationship.

**Simulated system**

The FFM tip was modeled by a capped (5,5) or (10,10) single-wall carbon nanotube (SWCNT), a spherical surface of a C540 fullerene, or a flat rigidly constrained sheet of graphene. The effective tip diameters for the (5, 5) and (10, 10) SWCNTs, and the C540 tip were respectively 1.2 nm, 2.2 nm, and 2.8 nm. The flat sheet tip had dimensions of 2 nm × 1.5 nm. For each tip-sample configuration, the feedback-controlled contact force $F_c$ was decreased from +10 nN to the negative pull-off force value. The value of $F_c$ is calculated as the total force exerted on the tip by the sample in the out of plane (Z-) direction, which, on average, is equal to the external applied load. The upper half of all the atoms in the simulated scanning tips were rigidly translated at a prescribed velocity of 5 nm/ns, while the lower half interacted freely with the sample. Monolayer suspended graphene samples (dimensions 11 nm × 12.5 nm) with periodic boundary conditions in the Y-direction and harmonically restrained boundaries in the X-direction were set up as shown in Fig. 1 (a). The scans were performed in the Y-direction (along the trench) in order to avoid any effect of the boundary in the direction of the scan. In addition, we simulated FFM scans of a 3-D sample with a fully supported base in order to compare with the results obtained for suspended monolayer graphene. The sample consisted of five 5.5 nm × 6.2 nm AB-stacked monolayers with all atoms in the lowest layer harmonically restrained against displacement [6]. The simulations were performed at T = 300 K, as described in more detail in Section 1 of the Supporting Information and the cited literature thereof.[14-16]

Figure 1(b) shows results from molecular statics simulations of lateral force versus lateral scan position for a 1.2 nm diameter SWCNT tip sliding on graphene. As expected from experiments and prior simulations, the data show a clear periodicity corresponding to the stick-slip behavior as the tip scans individual carbon atoms. Because of the static nature of the simulations, there is no average offset to the lateral force [17]. At higher contact forces $F_c$, a larger energy barrier exists for the tip to traverse an atom, resulting in the observed increasing stick-slip amplitude with increasing load. Note that the lateral force variations depend directly on the chiral direction of the scan [5], as well as the selected tip-sample interatomic interactions [17]. Depending on the problem, the van der Waals interactions may be better described by a more realistic many-body potential [18,19] than the pairwise Lennard-Jones potential used here. Since we consider lateral scans along a single chiral direction in absence of tip rotation, the use of Lennard-Jones tip-sample interaction is adequate and consistent with literature [4-6, 10, 13]. Moreover, as we show below, it reproduces the experimentally observed atomic detail of the sliding process in graphene.

**Results and discussion**

In order to further quantify friction and determine the effects of sample deformation in the form of the contribution from the local buckling deformation, or viscoelastic ploughing [20], dynamic simulations with energy dissipation were performed, as described earlier; resulting plots for the lateral force opposing sliding versus the lateral position are shown for a 1.2 nm diameter SWNT tip for three different contact forces in Fig. 1 (c). Despite the thermally induced irregularities, for the $F_c$ values of 0 nN and -0.75 nN the data generally corresponds to the static results in Fig. 1 (b) in terms of the amplitude and ~2.5 Å periodicity of the stick-slip events. For lower positive contact forces, our results are in fair agreement with previous experimental work [21] and reproduce experimentally observed atomic stick-slip in graphene [2] (see Fig. S1 (a) in the Supporting Information for stick-slip variations for various scanning tips). However, at $F_c = 5$ nN, the friction is strongly affected by a viscoelastic ploughing contribution where additional energy is spent to displace the viscoelastically behaving out of plane asperity [2,6]. The result is that the shown stick-slip behavior is significantly less regular than for data at lower loads that exhibited a smaller ploughing contribution (for comparison, see results for a supported multilayer sample with negligible ploughing in Fig. S1 (b) of Supporting Information). The lateral force variations shown in Fig. 1 (c) have a small nonzero average acting in the direction opposite the scan direction. This average value is the continuum-like friction force $F_f$ that arises from the atomistic contact and is the subject of our further discussion.

Shown in Fig. 2 (a) is a set of friction force $F_f$ versus contact force $F_c$ curves obtained for the four different tip configurations. In all cases, values of $F_c$ in the repulsive ($F_c > 0$) and adhesive ($F_c < 0$) regimes are shown. For $F_c > 0$ the friction force increases with increasing contact force, in agreement with existing tribological models. As $F_c$ becomes negative, the friction force on the 3D supported sample continues to decrease, maintaining a positive $\left(\frac{dF_f}{dF_c}\right)$ slope. The general trend is as expected for a single asperity contact with a load-dependent contact area, in agreement with well-established experimental knowledge [22-25]. However, as $F_c$ is reduced toward negative values on suspended graphene, the friction force *passes through a minimum value before increasing again*. At more negative loads there is a reduction in contact area, yet an increase in friction force up to the point of pull-off. This latter increase contradicts conventional tribological models and results in an anomalous $\left(\frac{dF_f}{dF_c}\right) < 0$, which corresponds to a system in which sliding *friction increases when two objects in contact are pulled away from one another*.

The observed phenomenon is considerably more pronounced for a different set of systems shown in Fig. 2 (b), in which we artificially increased the strength of the tip-sample van der Waals interaction four times, similar to what may happen experimentally in the presence of a weak capillary force. The enhanced interaction allows for significantly larger negative loads and corresponding Z-displacements of the surface compared to the systems in Fig. 2 (a).

In order to examine the underlying physical mechanism for the anomalous trend observed for $F_c < 0$, let us consider the energy dissipation mechanisms contributing to friction. The total opposing friction force $F_f$ is a sum of the van der Waals bonding-debonding process at the contact, and the ploughing component arising from lateral displacement of the deflected region shown in Figs. 3 (a) and (b) [26]. Thus,

$$F_f = F_{f,vdw} + F_{f,p},  \quad (1)$$

where $F_{f,vdw}$ and $F_{f,p}$ are the van der Waals and the ploughing contribution, respectively. Classical friction models for three-dimensional samples have been previously developed for $F_{f,vdw}$ [7, 8, 27] and $F_{f,p}$ [20, 28, 29]. Because suspended graphene behaves as a thin membrane [30] capable of significant deformation at $F_c > 0$ and $F_c < 0$, we derived a simple model that estimates the total amount of friction according to Eq. (1) (see sections 3-5 of Supporting Information for derivation). The resulting analytical $F_f$ vs. $F_c$ curves are shown in the insets of Figs. 2 (a, b), in good qualitative agreement with MD simulation results (see Supporting Information for the parameters used in the analysis). The observed behavior can be understood qualitatively from a close examination of the deformations of the graphene samples under the scanning tips.

Shown in Fig. 3 are the graphene deflection profiles from a cross-section taken at the center of the tip, along the scanning direction. The profiles are obtained from the snapshots of atomic positions during scanning. The tip-sample interaction strengths in Figs. 3 (a, c) and (b, d) correspond to the $F_f$ versus $F_c$ results shown in Figs. 2 (a) and (b), respectively. There is a clear qualitative similarity between the deformation profiles and thus the frictional contributions from ploughing in the adhesive and the repulsive mode shown respectively in Fig. 3 (a, c) and Fig. 3 (b, d). The energy is similarly spent on spatially displacing the deflected region in the direction of lateral scan, *regardless of the direction of out of plane deformation*. The main difference is that the deformed region during a scan at $F_c < 0$ is pulled laterally and upwards, as opposed to being pushed at $F_c > 0$. We therefore describe the regime where the apparent local friction-contact slope value is negative as *inverse ploughing*. The onset of its effect is determined by the applied force $F_c^0$ where $F_f$ is at minimum. The negative slope is then present for $F_c < F_c^0$ until pull-off because $F_{f,vdw}$ continues to decrease (due to decreasing effective contact area), while $F_{f,p}$ (which depends on the sample deformation energy – see section 5 of Supporting Information) increases with increasing deformation. The primary reason why this unusual trend is observable in graphene at $F_c < F_c^0$ is that the atomically thin sample is sufficiently compliant in the out-of-plane direction for the $F_{f,p}$ contribution to be comparable to $F_{f,vdw}$. Note that both $F_{f,p}$ and $F_{f,vdw}$ depend on the sample size (see Supporting Information), the scan velocity, as well as the local energy dissipation, as discussed in detail in [6] and the Supporting Information section thereof. The tip size dependence of the negative value of $\left(\frac{dF_f}{dF_c}\right)$ is attributed to a value of $F_{f,vdw}$ that increases with the tip radius, and a value of $F_{f,p}$ that is independent of tip size (assuming the tip radius is much smaller than the sample dimensions; see Eqs. S6 and S7 in the Supporting Information). Thus, *for a given negative load* that all tips can maintain without pull-off, a smaller tip will have a more negative $\left(\frac{dF_f}{dF_c}\right)$. However, a larger tip can ultimately induce larger stable negative loads than a smaller tip and may therefore have the largest negative value of the slope in the vicinity of pull-off. The flat rigid tip presents an extreme case where contact area is nearly constant and equal to the projected area of the plate. As a result, the contact area dependent $F_f$ contribution is also nearly constant with load, and all variations in $F_f$ can be ascribed to the load dependent ploughing contribution. The constant, relatively large contact area of the flat tip also allows for larger negative loads and higher absolute friction forces in the inverse ploughing regime. For all tip types it is observed that for a given load only one type of ploughing (conventional for $F_c > 0$ or inverse at $F_c < 0$) dominates. In principle, more complex, crater-like deformations that combine conventional and inverse ploughing may be possible with larger samples and larger radius of curvature tips. For the systems investigated here, neither the tight radii of the rounded tips, nor the flat tip allowed for these craters to form.

Another important point is that for all curves for spherical tips in Fig. 2, $F_f(F_c) > F_f(-F_c)$. The effect is likely due to the difference in the van der Waals component $F_{f,vdw}$ at equal deflections of opposite sign. The value $[F_f(F_c) - F_f(-F_c)]$ could in fact be a measure of the difference in the effective tip-sample contact area between the cases of ploughing and inverse ploughing.

**Conclusions**

In conclusion, we showed that the same viscoelastic deformation component that may be responsible for a layer-dependence of friction force for atomically thin layers can also produce an anomalous friction behavior in compliant graphene sheets at negative applied forces. Under adhesive pulling loads, the layer can be locally displaced upward, producing an asperity that must now be moved by a sliding FFM tip. The expenditure of energy to laterally displace this asperity can exceed the energy dissipation of traditional van der Waals friction, resulting in a case where friction force increases as the tip and sample are pulled apart. The findings of these simulations further the understanding of nanoscale friction, while providing insight for the design of novel nanoscale lubricants.

**Acknowledgements**

Part of this research was performed while A.S. held an NRC Postdoctoral Research Associateship at NIST. The authors are grateful to Rachel Cannara, Lauren Rast, David Read and Vinod Tewary for numerous illuminating discussions and valuable suggestions.

**Figures**

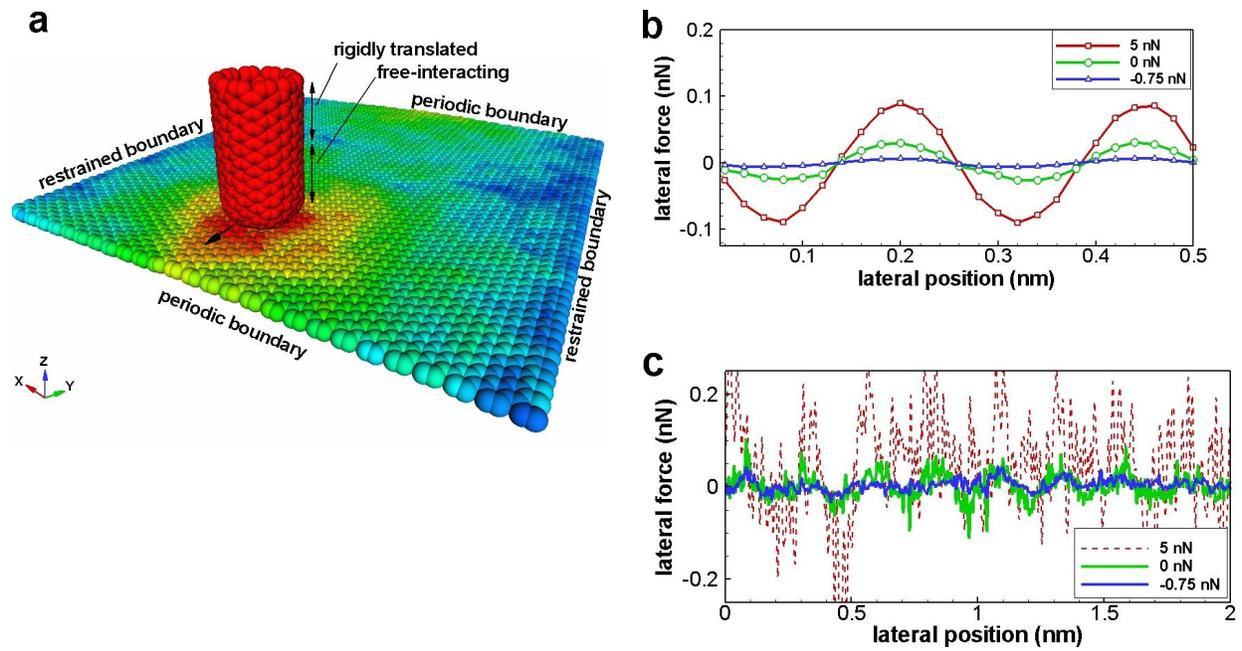

Figure 1. An example of the simulated atomistic system scanning in the adhesive (pulling) regime (a), atomically resolved variation of lateral force acting upon a 1.2 nm diameter tip from static (b) and dynamic simulations at 300 K (c).

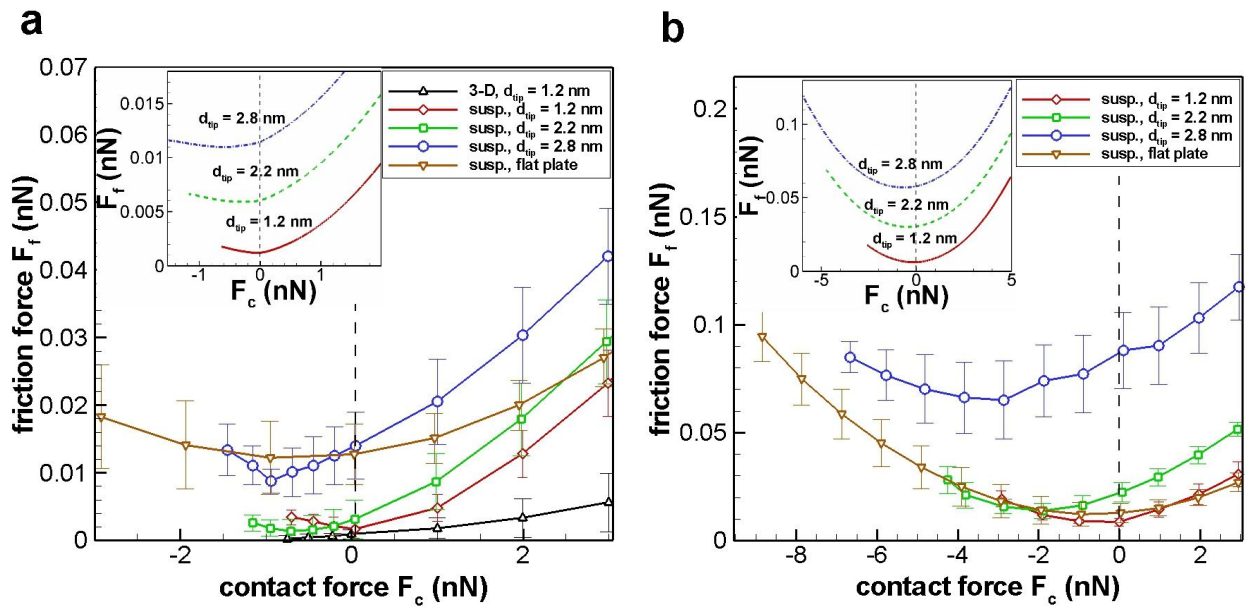

Figure 2. $F_f$ vs. $F_c$ curves (insets show analytical estimates for spherical tips) with (a) tip-sample empirical graphite interlayer adhesion, and (b) tip-sample interactions increased four times from (a). See Section 6 of Supporting Information for full-range force sweeps.

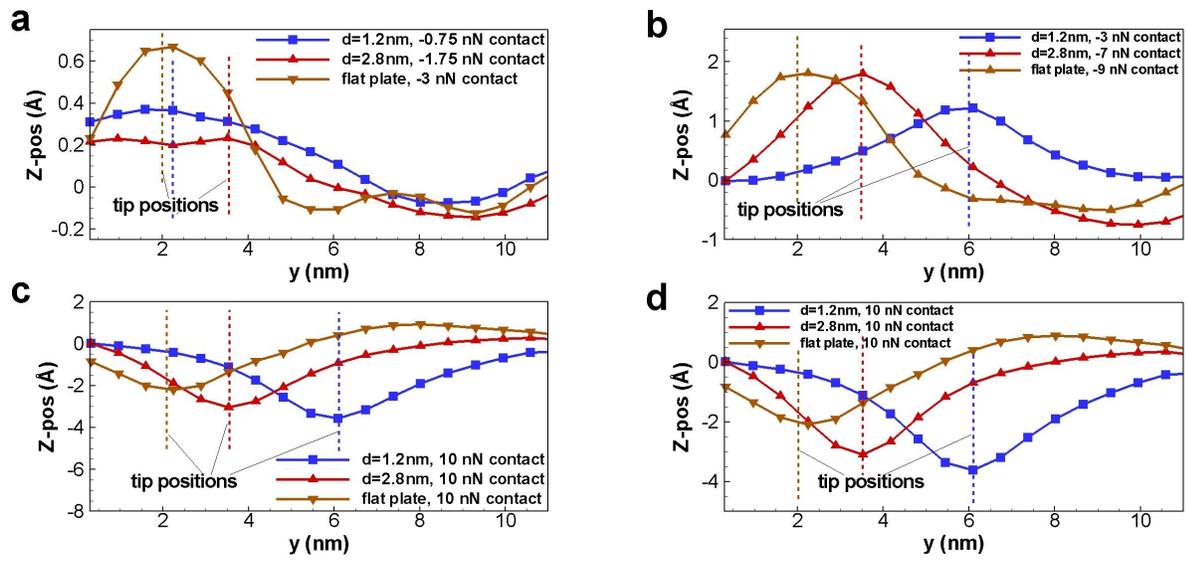

Figure 3. One-dimensional trace of the out-of-plane deformation of graphene samples during scanning in adhesive (a, b) and repulsive (c, d) modes.

**Supporting information for the paper:** *Anomalous friction in suspended graphene*
by A. Smolyanitsky and J.P. Killgore

*1. Molecular dynamics setup*

The carbon atoms within the sample and the tip interacted via the Tersoff-Brenner bond-order potential [1], while all tip-sample van der Waals interactions were modeled by a Lennard-Jones pair potential with $\varepsilon = 7.5$ meV and $\sigma = 3.1$ Å and an interaction cut-off distance of 7 Å. The atomic trajectories were integrated using the Verlet scheme with a time step of 1 fs. The calculated interlayer cohesion was 42.8 meV/atom at 3.35 Å of separation, in good agreement with experimental and *ab initio* data [2,3], addressed in the main text as "empirical". A temperature $T = 300$ K was maintained by use of a Langevin thermostat, as described in detail elsewhere [4]. All scans were performed during a simulated period of 1.0 ns, corresponding to a total scan distance of 5.0 nm. The force averages were calculated from the last 750 ps of each simulation, allowing 250 ps for relaxation. In addition, we simulated FFM scans of a 3-D sample with a fully supported base in order to compare with the results obtained for suspended monolayer graphene. The sample consisted of five 5.5 nm × 6.2 nm AB-stacked monolayers with all atoms in the lowest layer harmonically restrained against displacement [4].

*2. Dynamic stick-slip at 300K*

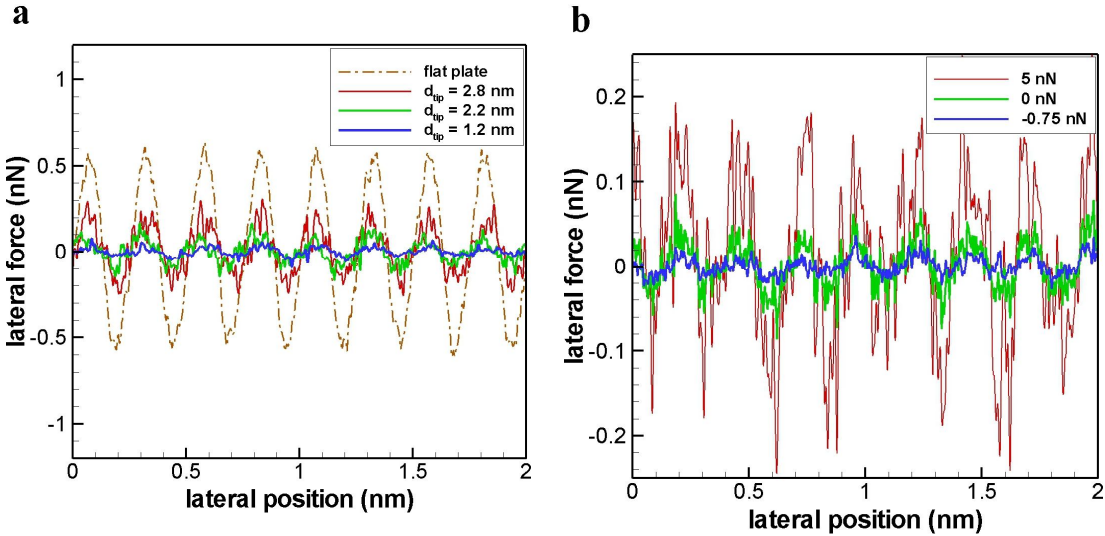

Figure. S1. Dynamic variation of the lateral force at $T = 300$ K for the suspended graphene samples scanned by various tips at a contact force of 0 nN (a) and for the supported 3-D sample scanned by a 1.2 nm diameter nanotube at different tip-sample contact force values (b).

*3. Dynamic friction between round tip and suspended circular membrane with viscoelastic ploughing*

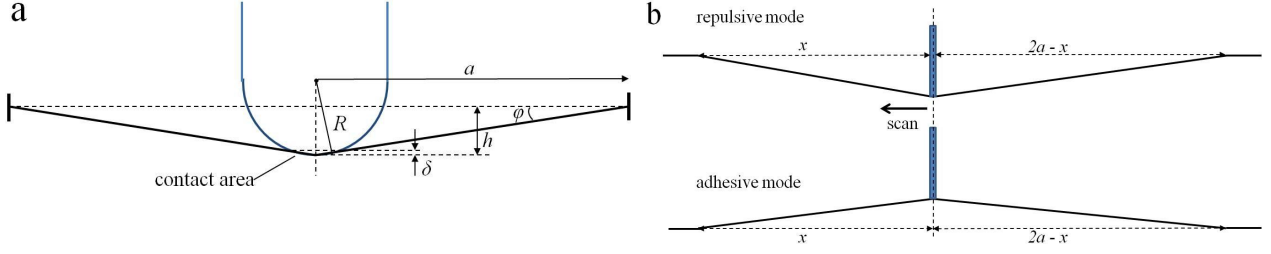

Figure S2. (a) Circular membrane with clamped edge deflected by a round tip. (b) Schematic representation of the repulsive and adhesive modes of a lateral scan.

We define the total friction force opposing the scanning tip as:

$$F_f = F_{f,vdw} + F_{f,p},\qquad \text{(S1)}$$

where $F_{f,vdw}$ and $F_{f,p}$ are the van der Waals and the ploughing contribution, respectively.

*4. Van der Waals contribution to friction*

$$F_{f,vdw} = \tau A,\qquad \text{(S2)}$$

where $\tau$ is the tip-sample interfacial shear strength. Unlike the conventional theories for a 3-D sample, here we have a thin elastic membrane with a negligible bending stiffness under a central load [5]. For our estimate, we assume a linear radial dependence of the out-of-plane deflection away from the center indented by a rigid tip. The effective contact area $A$ is given by

$$A = \pi R \delta,\qquad \text{(S3)}$$

where, by use of the dimensions shown in Fig S2 (a), for small $h > 0$:

$$\delta = R(1 - \cos\varphi) \approx \frac{Rh^2}{2a^2}.\qquad \text{(S4)}$$

The value of $h$ is obtained for a thin stretchable membrane without lateral pre-strain [5]:

$$h = \left(\frac{F_c a^2}{E^{2D}}\right)^{1/3},\qquad \text{(S5)}$$

where $E^{2D}$ is the second-order elastic stiffness of graphene, $F_c$ is the vertical load. We introduce the effect of intersurface cohesion by augmenting $F_c$ similarly to the JKR and DMT theories [6, 7]. From Eqs. (S3) and (S5) we then obtain an estimate of the contact area:

$$A = \frac{\pi R^2}{2a^2}\left(\frac{a^2(F_c + \varepsilon\pi R\gamma)}{E^{2D}}\right)^{2/3},\qquad \text{(S6)}$$

where $\gamma$ is the tip-sample cohesion energy and $\varepsilon$ is a dimensionless parameter selected to fit the pull-off force values obtained in simulation. Combined with Eq. (S2), an estimate of $F_{f,vdw}$ is obtained. Note that our expression, similarly to the JKR and DMT approaches, does not fully account for the "bulging" of the sample in the adhesive mode ($h < 0$), which may involve more complex crater-like deformations. However, we believe that the presented description is acceptable as a rough estimate.

*5. Ploughing (and inverse ploughing) contribution to friction*

The ploughing contribution can be estimated in a variety of ways. For a 3-D material, it has been described in relative detail [8]. For a thin membrane at small deflections, a much simpler estimate can be made. For $R \ll a$ and small out-of-plane deflections, the cases shown in Fig. S2 (b) are qualitatively similar, and the arising opposing force is due to asymmetry in the lateral component of the elastic forces at the contact. Ultimately, the shown deflection profiles are rough approximations of those shown in Fig. 3 of the main text.

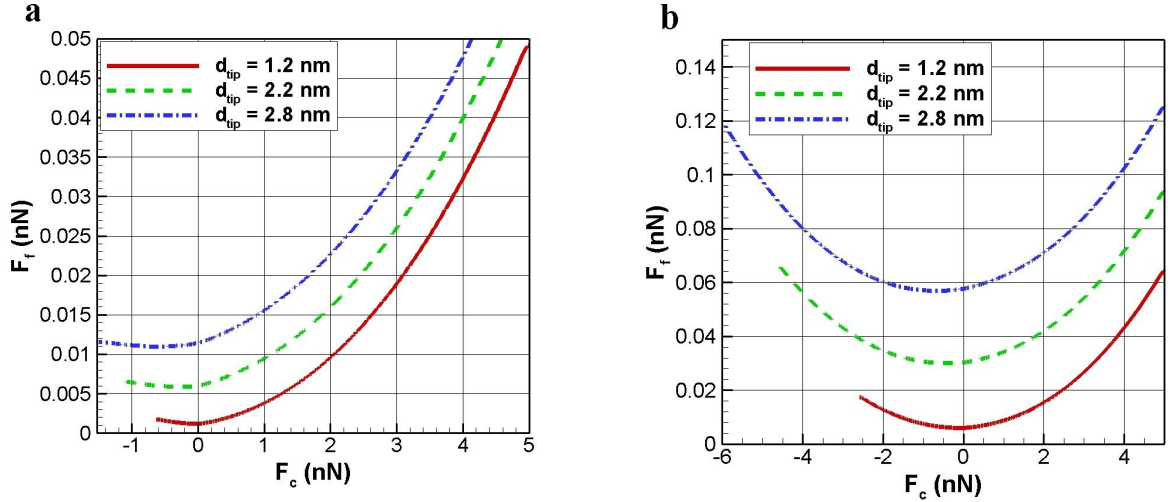

Figure S3. Analytically obtained $F_f$ vs $F_c$ curves shown as insets in Fig. 2 of the main text. (a) $\gamma = 42 \frac{meV}{atom}, \tau = 250\ MPa$. (b) $\gamma = 168 \frac{meV}{atom}, \tau = 500\ MPa$.

One must note that the asymmetry considered here is due to the viscoelastic response of the membrane and *is not an effect of the bou*ndary, similar to the asymmetry causing the ploughing effect in 3-D samples [8]. Moreover, one notes that the overall deflection given by Eq. (S5) corresponds to deflecting a one-dimensional elastic string with an elastic constant of the order of $E^{2D}$. Given that even in the case of a 2-D membrane the asymmetry is effectively one-dimensional in the direction of the scan, the total opposing force can then be estimated from the one-dimensional case as

$$F_{f,p} = -\left(\frac{dW}{dx}\right)_{x=a-\delta a} \approx \frac{\lambda E^{2D}(\delta a)}{a^4} h^4 = \frac{\delta a'}{(a^4 E^{2D})^{1/3}} F_c^{4/3}, \qquad (S7)$$

where $W = \int F_c(h)dh$ is the sample deformation energy, $\delta a \ll a$ is the effective measure of asymmetry (corresponding to the viscoelastic response of the sample and depending on the energy dissipation in the system, temperature, and scan velocity; see a representative example of asymmetry in the deflection profile obtained in Fig. S4) shown in Fig. S1 (b), and $\lambda$ is a dimensionless coefficient. We combine $\delta a' = \lambda \delta a$, which is now an independent fitting parameter. Note that the limiting case of $\delta a = 0$ corresponds to an ideally elastic response of the membrane (or a static tip-sample interaction), and the effect of ploughing is effectively eliminated, similar to the discussion presented elsewhere [8]. Also, one should keep in mind that $\delta a'$ is not constant for all values of $F_c$ and in general can vary depending on $h$ and $a$. We used a simple linear dependence on $F_c$, yielding $\delta a'$ values of the order of a few angstroms. The data presented in the insets of Fig. 2 (a) of the main text was obtained with $E^{2D} = 342 \frac{N}{m}$ [5], $\gamma = 42 \frac{meV}{atom} = 0.27 \frac{J}{m^2}$, $\tau = 250\ MPa$ [9], $\varepsilon = 4.0$, and circular samples and tips of effective dimensions used in our simulations. For Fig. 2 (b) of the main text, we used $\gamma_1 = 4\gamma$ and $\tau_1 \cong 2\tau$ [10]. The complete analytical curves from the insets of Fig. 2 of the main text are shown in Fig. S3. The observed discrepancies between simulation results and analytical estimates are due to the apparent simplicity of our analytical model, especially in terms of the boundary conditions and the assumed radial deformation profile. Also, the presented analytical model may be reaching its limit of applicability for larger tips, because with the tip radii of 1.1 nm and 1.4 nm, and $a \sim 6.5$ nm, the assumption $R \ll a$ is no longer valid.

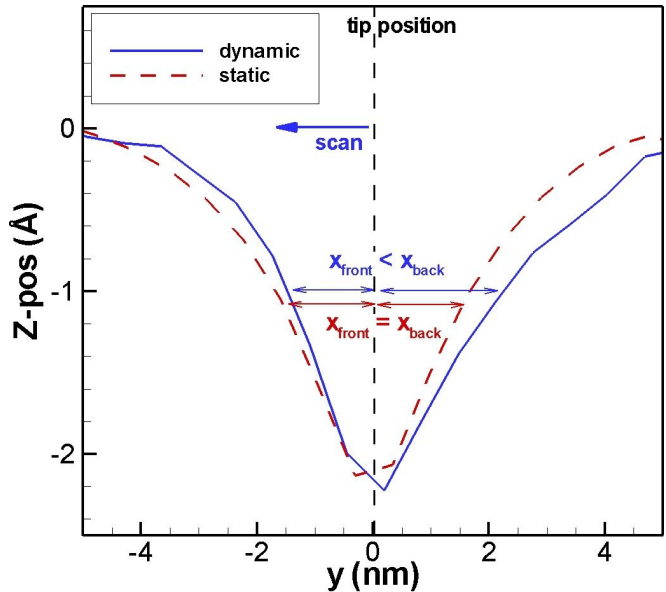

Figure S4. Asymmetric distribution of sample graphene directly from the atomic coordinates in a dynamic compared with symmetric distribution from a static simulation for a contact force of +5 nN, 1.2 nm diameter tip and *ab-initio* calculated tip-sample interaction. The data was obtained identically to that in Fig. 3 of main text. The plot was shifted so the tip is positioned at y = 0 for clarity.

*6. Full range friction-contact sweeps*

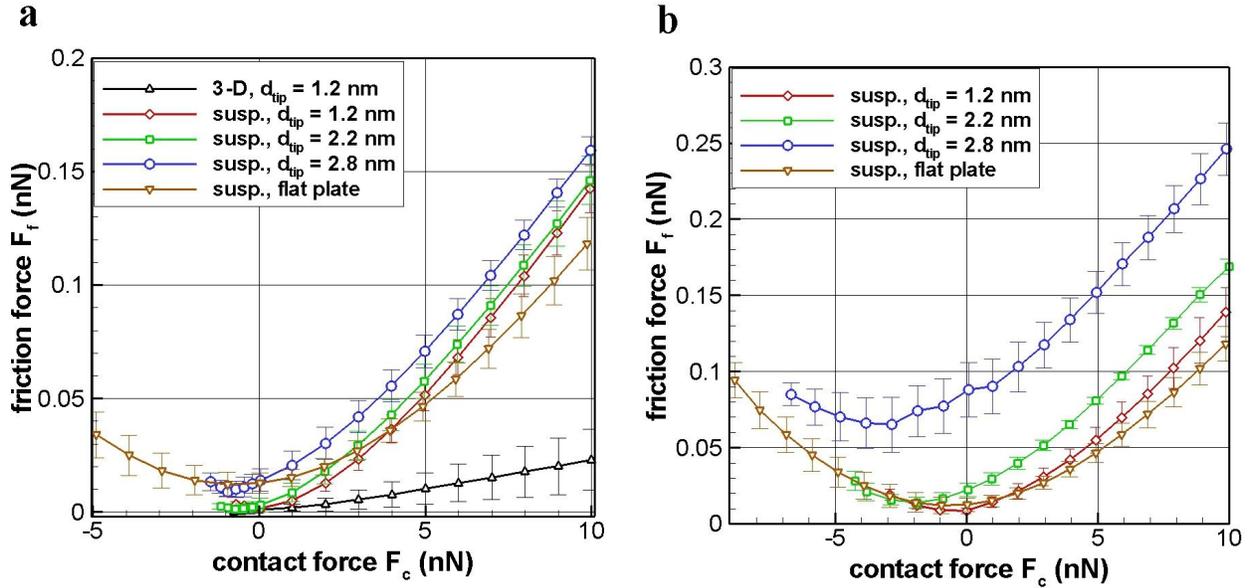

Figure S5. Friction force vs. full-range sweep of contact force (a) corresponding to Fig. 2 (a) of main text and (b) is corresponding to Fig. 2 (b) of main text.